\documentclass[preprint,authoryear,times,12pt]{elsarticle}

\usepackage{lipsum}
\makeatletter
\def\ps@pprintTitle{%
 \let\@oddhead\@empty
 \let\@evenhead\@empty
 \def\@oddfoot{}%
 \let\@evenfoot\@oddfoot}
\makeatother

\usepackage{ulem} 

\usepackage[dvipsnames,table]{xcolor}

\definecolor{blue}{RGB}{0, 0, 225}

\usepackage[labelfont=bf,textfont=md]{caption}

\usepackage[hidelinks]{hyperref} 

\usepackage[a4paper,margin=1in]{geometry}

\usepackage[super]{nth}

\usepackage{eurosym} 

\usepackage{multirow}

\usepackage{hhline}

\usepackage{longtable}

\usepackage{lscape} 

\usepackage[toc,page]{appendix}

\usepackage{float} 

\usepackage{mathtools} 
\usepackage{setspace}

\usepackage{array}
\newcommand{\PreserveBackslash}[1]{\let\temp=\\#1\let\\=\temp}
\newcolumntype{C}[1]{>{\PreserveBackslash\centering}p{#1}}
\newcolumntype{R}[1]{>{\PreserveBackslash\raggedleft}p{#1}}
\newcolumntype{L}[1]{>{\PreserveBackslash\raggedright}p{#1}}

\begin{document}

\title{Financial statements of companies in Norway\tnoteref{t1}}
\tnotetext[t1]{I appreciate good support from the Br\o nn\o ysund Register Centre (Norwegian: "Registerenheten i Brønnøysund"). This work acknowledges financial support from NTNU Business School, Faculty of Economics and Management at the Norwegian University of Science and Technology.}
\author[1]{Ranik Raaen Wahlstrøm\corref{cor1}}

\cortext[cor1]{E-mail: \href{mailto:ranik.raaen.wahlstrom@ntnu.no}{ranik.raaen.wahlstrom@ntnu.no} \hspace{0.2cm} Web page: \href{https://www.ntnu.edu/employees/ranik.raaen.wahlstrom}{www.ntnu.edu/employees/ranik.raaen.wahlstrom}}
\address[1]{NTNU Business School, Norwegian University of Science and Technology, 7491 Trondheim, Norway}

\begin{abstract}
This document details a dataset that contains all unconsolidated annual financial statements of the universe of Norwegian private and public limited liability companies. It also includes all financial statements of other company types reported to the Norwegian authorities.

\end{abstract}


\maketitle



\doublespacing

\section{The dataset}

I compile and harmonize data from various providers, including the Br\o nn\o ysund Register Centre, about Norwegian companies in different formats. This data is transformed into a comprehensive and manageable two-dimensional dataset, stored in comma-separated values (CSV) files for easy use in popular software and programming languages such as Excel, Python, R, Matlab, and STATA.
The dataset comprises all unconsolidated annual financial statements of Norwegian companies reported to the Norwegian authorities for the accounting years 2006-2023. Reporting these statements is mandatory for all Norwegian private and public limited liability companies. Therefore, the dataset includes all unconsolidated annual financial statements from these companies. Additionally, it contains the financial statements of other types of companies reported to the authorities, such as sole proprietorships and general partnerships.\footnote{The reporting of financial statements to the authorities for companies that are not limited liability companies is voluntary or only required in certain cases, e.g., for companies of a given size.}
In total, the dataset contains 5,470,982 financial statements. Table \ref{tab:orgform} presents the number of financial statements in the dataset per accounting year and organizational form.\footnote{The organizational form of the financial statements’ underlying companies are given in the dataset by the column \textit{"orgform"}. A complete list with descriptions of the organizational forms in Norway is available at \href{https://www.brreg.no/bedrift/organisasjonsformer/}{www.brreg.no/bedrift/organisasjonsformer}}

\begin{table}[h]
\begin{center}
\captionof{table}{\textbf{Financial statements per accounting year and organizational form}}\label{tab:orgform}
    \begin{tabular}{r|rrrrrr|r}
\hline\hline
          & \multicolumn{1}{c}{\textbf{AS}} & \multicolumn{1}{c}{\textbf{ASA}} & \multicolumn{1}{c}{\textbf{ENK}} & \multicolumn{1}{c}{\textbf{ANS}} & \multicolumn{1}{c}{\textbf{DA}} & \multicolumn{1}{c|}{\textbf{Other}} & \multicolumn{1}{c}{\textbf{Total}} \\ \hline
    \textbf{2006} & 181,402 & 246   & 3,202 & 1,504 & 819   & 22,525 & 209,698 \\
    \textbf{2007} & 193,859 & 257   & 3,028 & 1,473 & 853   & 24,646 & 224,116 \\
    \textbf{2008} & 200,520 & 250   & 3,760 & 2,137 & 1,166 & 28,904 & 236,737 \\
    \textbf{2009} & 202,208 & 242   & 3,649 & 2,111 & 1,512 & 30,529 & 240,251 \\
    \textbf{2010} & 204,564 & 244   & 3,492 & 2,019 & 1,506 & 31,787 & 243,612 \\
    \textbf{2011} & 210,286 & 239   & 3,372 & 1,951 & 1,500 & 32,619 & 249,967 \\
    \textbf{2012} & 225,121 & 202   & 3,281 & 1,884 & 1,497 & 32,288 & 264,273 \\
    \textbf{2013} & 238,808 & 197   & 3,118 & 1,818 & 1,415 & 31,254 & 276,610 \\
    \textbf{2014} & 252,215 & 195   & 3,009 & 1,769 & 1,352 & 30,154 & 288,694 \\
    \textbf{2015} & 265,555 & 189   & 2,824 & 1,722 & 1,340 & 29,997 & 301,627 \\
    \textbf{2016} & 279,718 & 185   & 2,817 & 1,536 & 1,304 & 30,030 & 315,590 \\
    \textbf{2017} & 294,366 & 182   & 2,705 & 1,365 & 1,265 & 29,957 & 329,840 \\
    \textbf{2018} & 308,318 & 180   & 2,668 & 1,218 & 1,251 & 29,999 & 343,634 \\
    \textbf{2019} & 321,049 & 183   & 2,573 & 1,142 & 1,182 & 30,249 & 356,378 \\
    \textbf{2020} & 336,504 & 205   & 2,345 & 1,071 & 1,115 & 30,357 & 371,597 \\
    \textbf{2021} & 358,181 & 216   & 2,201 & 1,015 & 1,077 & 30,595 & 393,285 \\
    \textbf{2022} & 372,303 & 205   & 2,018 & 947   & 1,060 & 31,124 & 407,657 \\
    \textbf{2023} & 384,871 & 201   & 1,962 & 940   & 1,061 & 31,750 & 420,785 \\
    \textbf{2024} & 391,256 & 197   & 1,732 & 838   & 1,017 & 31,671 & 426,711 \\
    \hline
    \textbf{Total} & 5,221,104 & 4,015 & 53,756 & 28,460 & 23,292 & 570,435 & 5,901,062 \\
\hline\hline
\end{tabular}
\caption*{AS is the abbreviation for private limited liability companies (i.e., not listed). ASA stands for public limited liability companies (i.e., listed). ENK is the abbreviation for sole proprietorships, while ANS and DA represent general partnerships. A comprehensive list of organizational forms included in the dataset can be found at \href{https://www.brreg.no/bedrift/organisasjonsformer/}{www.brreg.no/bedrift/organisasjonsformer}.}
    \end{center}
\end{table}

Each row in the dataset represents a firm-year, i.e., an annual financial statement of a company for a specific accounting year.\footnote{Some companies have submitted the financial statement for the same accounting year multiple times to the Norwegian authorities, typically due to error corrections. In such cases, only the last submitted financial statement is included in the dataset.}
For each of these firm-years (rows), several accounting items (columns) are available from both the balance sheet and income statement, as detailed in \ref{appendix:columnsFinancial}. There are no missing values for the financial information in \ref{appendix:columnsFinancial}. If a value is missing in the dataset for any of the items listed in \ref{appendix:columnsFinancial}, it indicates that no value was provided when the financial statement was reported, that is, the value is zero.

To ensure comparability of the financial statements, all values of the accounting items listed in \ref{appendix:columnsFinancial} have been converted to Norwegian krone (NOK) for all financial statements.\footnote{For financial statements reported in currencies other than NOK, their financial data is converted to NOK using the daily exchange rate on their balance sheet date. If there is no exchange rate available for the balance sheet date due to weekends or public holidays, the last available exchange rate prior to the balance sheet date is used. All exchange rates are retrieved from Norges Bank at \href{https://www.norges-bank.no/en/topics/Statistics/exchange_rates}{www.norges-bank.no/en/topics/Statistics/exchange\_rates}}

As all columns listed in \ref{appendix:columnsFinancial} are in Norwegian, an English translation of some of the commonly used accounting items in the literature is provided in \ref{appendix:commonAccountingItems}.

The dataset also includes other information about the financial statements and their underlying companies, including two columns indicating total assets and total turnover, respectively, in EUR (columns \textit{"sum\_omsetning\_EUR"} and \textit{"sum\_eiendeler\_EUR"}). An overview is provided in \ref{appendix:columnsNonFinancial}. Additionally, audit information per financial statement is detailed in \ref{appendix:columnsAuditor}.

\section{Code examples}

Examples of Python code for managing the dataset can be found in a GitHub repository at the following link: \href{https://FinancialStatementsNorway.ranik.no}{https://FinancialStatementsNorway.ranik.no}

\section{Research outcome}

The dataset has proven to be an invaluable resource for research. This includes the research mentioned in the following non-exhaustive list:

\vspace{-0.3cm}
\begin{itemize}
\setlength\itemsep{-0.1cm}
\item \cite{paraschiv_bankruptcy_2026}
\item \cite{berg_kostnadsasymmetri_2024}
\item \cite{wahlstrom_enhancing_2024}
\item \cite{opstad_longrun_2022}
\item \cite{opstad_nonlinear_2022}
\item \cite{opstad_profit_2023}
\item \cite{kainth_IFRS_2021}
\item \cite{pelja_hvordan_2021}
\end{itemize}

\subsection*{PhD Theses}
\vspace{-0.3cm}
\begin{itemize}
\setlength\itemsep{-0.1cm}
\item \cite{nilsen_accounting_2026}
\item \cite{kainth_essays_2023}
\item \cite{wahlstrom_financial_2021}
\end{itemize}

\subsection*{MSc Theses}
\vspace{-0.3cm}
\begin{itemize}
\setlength\itemsep{-0.1cm}
\item \cite{fiskvik_prediksjon_2026}
\item \cite{dyrseth_overlevelse_2026}
\item \cite{amalieronning_kjonnsmangfold_2025}
\item \cite{biskopshavn_resultatjustering_2025}
\item \cite{ivanlarsen_early_2025}
\item \cite{nicolascaccamolsen_enhancing_2025}
\item \cite{lyso_predicting_2024}
\item \cite{overland_lonnsomhetsdrivere_2024}
\item \cite{strand_konkursprediksjon_2024}
\item \cite{vareide_resultatjustering_2024}
\item \cite{dahl_konkursprediksjon_2023}
\item \cite{fjerstad_prediksjon_2023}
\item \cite{hakenstad_konkursprediksjon_2023}
\item \cite{nygard_earnings_2023}
\item \cite{odegard_explainable_2023}
\item \cite{becker_ikkefinansiell_2022}
\item \cite{brondbo_konkursprediksjon_2022}
\item \cite{hagen_revisorskifte_2022}
\item \cite{kaspersen_interpretable_2022}
\item \cite{lauvsnes_konkursprediksjon_2022}
\item \cite{nymoen_real_2022}
\item \cite{moen_bankruptcy_2020}
\end{itemize}

\clearpage
\singlespacing
\bibliographystyle{elsarticle-harv-nourldoi}
\bibliography{bibliography}
\doublespacing

\clearpage
\begin{appendices}
\appendix

\setcounter{table}{0}
\section{Available financial information}
\label{appendix:columnsFinancial}

All accounting items available in the data per financial statement are listed in Table \ref{tab:columns:financial}. They are presented in Norwegian. An English translation of some of the commonly used accounting items in the literature is provided in \ref{appendix:commonAccountingItems}.

The items \textit{"Udekket tap"}, \textit{"Minoritetsinteresser"}, and \textit{"Utbytte"} are provided by the data provider for both the financial statements’ balance sheets and income statements. However, in the dataset, it is the balance sheet values of \textit{"Udekket tap"} and \textit{"Minoritetsinteresser"} as well as the income statement values of \textit{"Utbytte"} that are included.

\vspace{0.5cm}

\captionof{table}{\textbf{Available accounting items for each financial statement sorted alphabetically}}\label{tab:columns:financial}
\singlespacing
\begin{longtable}{l}
\textbf{Accounting item}\\
\hline
\endfirsthead
\caption*{\centering \textbf{Table \ref{tab:columns:financial}: Available accounting items for each financial statement sorted alphabetically (continued)}}\\
\textbf{Accounting item}\\
\hline
\endhead
\endfoot
\hline
    Aksjer og andeler i foretak i samme konsern \\
    Andre avsetninger for forpliktelser \\
    Andre driftskostnader (fou kostnader, adm.kostnader, markedsfoeringskostnader) \\
    Andre finansielle instrumenter \\
    Andre fordringer \\
    Andre immaterielle eiendeler \\
    Andre kortsiktige fordringer \\
    Andre markedsbaserte finansielle instrumenter \\
    Andre resultatkomponenter for IFRS-foretak \\
    Annen driftsinntekt \\
    Annen driftskostnad \\
    Annen egenkapital \\
    Annen Egenkapital \\
    Annen finansinntekt \\
    Annen finanskostnad \\
    Annen innskutt egenkapital \\
    Annen kortsiktig gjeld \\
    Annen renteinntekt \\
    Annen rentekostnad \\
    Ansvarlig laanekapital \\
    Avgitt konsernbidrag \\
    Avsatt utbytte \\
    Avskrivning paa varige driftsmidler og immaterielle eiendeler \\
    Bankinnskudd, kontanter og lignende \\
    Beholdning av egne aksjer \\
    Betalbar skatt \\
    Biologiske eiendeler \\
    Driftsloesoere, inventar, verktoey, kontormaskiner og lignende \\
    Driftsresultat \\
    Ekstraordinaer inntekt \\
    Ekstraordinaer kostnad \\
    Ekstraordinaere poster \\
    Ekstraordinaert utbytte \\
    Endring i beholdning av egentilvirkede anleggsmidler \\
    Endring i beholdning av varer under tilvirkning og ferdig tilvirkede varer \\
    Fond \\
    Fond for urealiserte gevinster \\
    Fond for vurderingsforskjeller \\
    Fondsemisjon \\
    Forskning og utvikling \\
    Garantistillelser \\
    Gjeld til kredittinstitusjoner \\
    Goodwill \\
    Ikke registrert kapitalforhoeyelse \\
    Inntekt paa andre investeringer \\
    Inntekt paa investering i annet foretak i samme konsern \\
    Inntekt paa investering i datterselskap \\
    Inntekt paa investering i datterselskap og tilknyttet selskap \\
    Inntekt paa investering i tilknyttet selskap \\
    Investering i annet foretak i samme konsern \\
    Investering i datterselskap \\
    Investeringer i aksjer og andeler \\
    Investeringer i tilknyttet selskap \\
    Investeringseiendom \\
    Kasse/Bank/Post \\
    Konserfordringer \\
    Konsernbidrag \\
    Konsernfordringer \\
    Konsesjoner, patenter, lisenser, varemerker og lignende rettigheter \\
    Konvertible laan \\
    Kortsiktig konserngjeld \\
    Krav paa innbetaling av selskapskapital \\
    Kundefordringer \\
    Langsiktig konserngjeld \\
    Leverandoergjeld \\
    Loennskostnad \\
    Loennskostnader \\
    Laan til foretak i samme konsern \\
    Laan til tilknyttet selskap og felles kontrollert virksomhet \\
    Markedsbaserte aksjer \\
    Markedsbaserte obligasjoner \\
    Maskiner og anlegg \\
    Minoritetsinteresser \\
    Nedskrivning av andre finansielle omloepsmidler \\
    Nedskrivning av finansielle anleggsmidler \\
    Nedskrivning av finansielle eiendeler \\
    Nedskrivning av varige driftsmidler og immaterielle eiendeler \\
    Netto finans \\
    Obligasjoner \\
    Obligasjoner og andre fordringer \\
    Obligasjoner og andre langsiktige fordringer \\
    Obligasjonslaan \\
    Oevrig langsiktig gjeld \\
    Ordinaert resultat \\
    Ordinaert resultat etter skattekostnad \\
    Ordinaert resultat foer skattekostnad \\
    Ordinaert utbytte \\
    Overfoering til/fra fond \\
    Overfoering til/fra fond for urealiserte gevinster \\
    Overfoering til/fra fond for vurderingsforskjeller \\
    Overfoeringer til/fra annen egenkapital \\
    Overkurs \\
    Overkursfond \\
    Pantstillelser \\
    Pensjonsforpliktelser \\
    Renteinntekt fra foretak i samme konsern \\
    Rentekostnad til foretak i samme konsern \\
    Resultat av ekstraordinaere poster \\
    Salgsinntekt \\
    Salgsinntekter \\
    Selskapskapital \\
    Sertifikatlaan \\
    Skattekostnad paa ekstraordinaere poster \\
    Skattekostnad paa ekstraordinaert resultat \\
    Skattekostnad paa ordinaert resultat \\
    Skip, rigger, fly og lignende \\
    Skyldige offentlige avgifter \\
    Sum anleggsmidler \\
    Sum annen langsiktig gjeld \\
    Sum avsetninger for forpliktelser \\
    Sum bankinnskudd, kontanter og lignende \\
    Sum driftsinntekter \\
    Sum driftskostnader \\
    Sum egenkapital \\
    Sum egenkapital og gjeld \\
    SUM EGENKAPITAL OG GJELD \\
    Sum eiendeler \\
    SUM EIENDELER \\
    Sum finansielle anleggsmidler \\
    Sum finansinntekter \\
    Sum finanskostnader \\
    Sum fordringer \\
    Sum gjeld \\
    SUM GJELD OG EGENKAPITAL \\
    Sum immaterielle eiendeler \\
    Sum innskutt egenkapital \\
    Sum inntekter \\
    Sum investeringer \\
    Sum kortsiktig gjeld \\
    Sum kostnader \\
    Sum langsiktig gjeld \\
    Sum omloepsmidler \\
    Sum opptjent egenkapital \\
    Sum overfoeringer og disponeringer \\
    Sum resultatkomponenter for IFRS-foretak \\
    Sum varer \\
    Sum varige driftsmidler \\
    Tilleggsutbytte \\
    Tomter, bygninger og annen fast eiendom \\
    Totalresultat \\
    Udekket tap \\
    Utbytte \\
    Utsatt skatt \\
    Utsatt skattefordel \\
    Varekostnad \\
    Varer \\
    Verdiendring av markedsbaserte finansielle omloepsmidler \\
    Verdioekning andre finansielle instrumenter vurdert til virkelig verdi \\
    Verdioekning av markedsbaserte finansielle omloepsmidler \\
    Verdireduksjon andre finansielle instrumenter vurdert til virkelig verdi \\
    Verdireduksjon av markedsbaserte finansielle omloepsmidler \\
    Aarsresultat \\
    Aarsresultat etter minoritetsinteresser \\
    Aarsresultat foer minoritetsinteresser \\
\end{longtable}

\clearpage
\setcounter{table}{0}
\section{English translation of accounting items commonly used in the literature}
\label{appendix:commonAccountingItems}

\begin{table}[h]
\begin{center}
\captionof{table}{\textbf{English translation of accounting items from the income statement}}\label{tab:common:income_statement}
    \begin{tabular}{l|l}
    \hline\hline
    \textbf{English translation} & \textbf{Norwegian as stated in the dataset} \\ \hline
    Total turnover & Sum inntekter \\
    Sales income & Salgsinntekt \\
    Salary costs & Loennskostnad \\
    Depreciations & Avskrivning på varige driftsmidler og immaterielle eiendeler \\
    Write-downs & Nedskrivning av varige driftsmidler og immaterielle eiendeler \\
    Interest income & Sum finansinntekter \\
    Interest expenses & Sum finanskostnader \\
    Income before tax & Ordinaert resultat foer skattekostnad \\
    Income after tax & Ordinaert resultat etter skattekostnad \\
    Net income & Aarsresultat \\
    Dividends & Utbytte \\
\hline\hline
\end{tabular}
    \end{center}
\end{table}

\begin{table}[h]
\begin{center}
\captionof{table}{\textbf{English translation of accounting items from the balance sheet}}\label{tab:common:balance_sheet}
    \begin{tabular}{l|l}
    \hline\hline
    \textbf{English translation} & \textbf{Norwegian as stated in the dataset} \\ \hline
    Total assets & SUM EIENDELER \\
    Current assets & Sum omloepsmidler \\
    Accounts receivable & Kundefordringer \\
    Book equity (total) & Sum egenkapital \\
    Paid-in equity & Sum innskutt egenkapital \\
    Retained earnings & Sum opptjent egenkapital \\
    Total debt & Sum gjeld \\
    Current liabilities & Sum kortsiktig gjeld \\
    Trade accounts payable  & Leverandoergjeld \\
    Public taxes payable & Skyldige offentlige avgifter \\
    Interest-bearing debt & Gjeld til kredittinstitusjoner \\
    Investments in subsidiaries & Investering i datterselskap \\
    Intangible assets & Sum immaterielle eiendeler \\
\hline\hline
\end{tabular}
    \end{center}
\end{table}

\clearpage
\setcounter{table}{0}
\section{Available other information}
\label{appendix:columnsNonFinancial}

\captionof{table}{\textbf{Available other information about the financial statements and their underlying companies}}\label{tab:columns:nonfinancial}
\singlespacing
\begin{longtable}{L{4.2cm}|L{10.8cm}}
\textbf{Name} & \textbf{Description} \\
\hline
\endfirsthead
\caption*{\centering \textbf{Table \ref{tab:columns:nonfinancial}: Available other information about the financial statements and their underlying companies (continued)}}\\
\textbf{Name} & \textbf{Description} \\
\hline
\endhead
\endfoot
\hline
    regnaar & Accounting year of the financial statement \\
    startdato & Starting date for the financial statement's financial year \\
    avslutningsdato & Balance sheet date and the end date for the financial statement's accounting year \\
    sum\_eiendeler\_EUR & Total assets in EUR \\
    sum\_omsetning\_EUR & Total turnover in EUR \\
    orgnr & Legal organization number for the underlying company \\
    orgform & Organizational form. An overview of organizational forms for Norwegian companies is available at https://www.brreg.no/bedrift/organisasjonsformer/ \\
    konkursdato & The date of filing for bankruptcy in the case the underlying company is filed for bankruptcy by itself or its creditors through a court. Missing value indicates that no filing for bankruptcy has occurred. All bankruptcy filings up to February 2026 are included. \\
    stiftelsesdato & The date of incorporation of the underlying company \\
    landkode & The country of the business address of the underlying company \\
    postnummer & The postal code of the business address of the underlying company \\
    age\_in\_days & The age of the underlying company in days at the reporting of the financial statement. This is given by the difference between the balance sheet date of the financial statement (avslutningsdato) and the date of incorporation of the underlying company (stiftelsesdato) \\
    naeringskode & Industry code for the underlying company in accordance with Standard Industrial Classification 2007 (SIC 2007). 00.000 indicates that the company is a holding company. For an overview, see \href{https://www.ssb.no/klass/klassifikasjoner/6}{www.ssb.no/klass/klassifikasjoner/6} \\
    naeringskoder\_level\_1 & Industry code level 1 for the underlying company in accordance with Standard Industrial Classification 2007 (SIC 2007)  \\
    naeringskoder\_level\_2 & Industry code level 2 for the underlying company in accordance with Standard Industrial Classification 2007 (SIC 2007). This is the same as the first two numbers of the column naeringskode \\
    mor\_i\_konsern & Dummy: one if the company is the mother of a company group \\
    regler\_smaa & Dummy: one if the financial statement follows accounting rules for small enterprises \\
    ifrs\_selskap & Dummy: one if the financial statement follows IFRS or simplified IFRS, zero if it follows Norwegian GAAP  \\
    mottakstype & Dummy: one if the financial statement was submitted electronically via the online platform Altinn, zero if it was registered manually \\
    avviklingsregnskap & Dummy: one if the financial statement is the last financial statement before liquidation of the underlying company \\
    feilvaloer & Dummy: one if the company has reported the figures in the wrong value, e.g., if it appears in the submitted financial statement that the company has reported figures in the whole thousand, but the accounting figures that appear in the result and balance sheet are not \\
    journalnr & The record number of the financial statement at the Brønnøysund Register Centre  \\
    fravalg\_revisjon & Dummy: one if the company has opted-out on external auditing \\
    utarbeidet\_regnskapsforer & Dummy: one if the financial statement is prepared by a certified public accountant  \\
    bistand\_regnskapsforer & Dummy: one if the financial statement is prepared with the assistance of a certified public accountant \\
    bankrupt\_fs &  Dummy: one if the financial statement is the last financial statement of the underlying company and the company has filed for bankruptcy in accordance with the variable \textit{konkursdato} (see above). Note that for \textit{konkursdato}, only bankruptcy filings up to February 2026 are included. \\
\end{longtable}

\clearpage
\setcounter{table}{0}
\section{Available audit information}
\label{appendix:columnsAuditor}

\captionof{table}{\textbf{Available audit information about the financial statements}}\label{tab:columns:auditor}
\singlespacing
\begin{longtable}{L{5.5cm}|L{9.5cm}}
\textbf{Name} & \textbf{Description} \\
\hline
\endfirsthead
\caption*{\centering \textbf{Table \ref{tab:columns:auditor}: Available auditor information about the financial statements (continued)}}\\
\textbf{Name} & \textbf{Description} \\
\hline
\endhead
\endfoot
\hline
    Revisornr & Legal organization number for the auditor company \\
    Revisor & Name of the auditor company \\
    Revisjonsanmerkninger &  Dummy: one if there are any audit remarks for the financial statement \\
    Revisors kommentar &  Auditor's comment \\
    Revisors presiseringer &  Auditor's clarifications \\
    Revisors presisering (tekstlig) &  Auditor's clarifications - textual \\
    Revisors forbehold &  Auditor's reservations \\
    Revisjonshonorar &  Audit fees \\
    Annet revisjonshonorar &  Other audit fees \\
\end{longtable}

\vspace{1cm}

As a general rule, public and non-public limited liability companies in Norway (AS and ASA) must have their financial statements audited by an external auditor. 
However, non-public limited liability companies (AS) can opt out of auditing of financial statements of 2011 and later if the following is true for the financial statement the \textit{previous} accounting year: total turnover is below NOK 5 millions, total assets is below NOK 20 millions, and average number of employees do not exceed ten man-years. For financial statements from 2017 and later, the two first thresholds (for the previous financial statement, i.e., of 2016 and later) were raised to NOK 6 millions and NOK 23 millions, respectively. Further, newly established on-public limited liability companies (AS) can opt out of auditing of their first financial statement if \textit{i)} the first financial statement is for 2011 or later, \textit{ii)} the company has no more than ten employees, and \textit{iii)} the opening balance is less than NOK 20 millions. The latter threshold was raised to NOK 23 millions for first financial statement of 2017 or later. 
Further, companies from the following industries can never opt out of auditing: 47.730 (\textit{'Dispensing chemist in specialised stores'}), 69.100 (\textit{'Legal activities'}), and 92.000 (\textit{'Gambling and betting activities'}).

\end{appendices}

\end{document}